**Evolution of entanglement within classical light states.**

R. Mark Stevenson[1], Andrew J. Hudson[1,2], Anthony J. Bennett[1], Robert J. Young[1], Christine A. Nicoll[2], David A. Ritchie[2], and Andrew J. Shields[1].

[1]Toshiba Research Europe Limited, 208 Cambridge Science Park, Cambridge CB4 0GZ, UK.

[2]Cavendish Laboratory, University of Cambridge, JJ Thompson Avenue, Cambridge CB3 0HE, UK

**Abstract**

We investigate the evolution of quantum correlations over the lifetime of a multi-photon state. Measurements reveal time-dependent oscillations of the entanglement fidelity for photon pairs created by a single semiconductor quantum dot. The oscillations are attributed to the phase acquired in the intermediate, non-degenerate, exciton-photon state and are consistent with simulations. We conclude that emission of photon pairs by a typical quantum dot with finite polarisation splitting is in fact entangled in a time-evolving state, and not classically correlated as previously regarded.

Entangled optical states provide fundamental insights into the nature of quantum mechanics, and are an essential resource for advanced quantum information applications such as scalable linear optics quantum computing[1] and quantum key distribution over large distances[2]. The realisation of entangled light sources has thus far concentrated on the time-averaged relationship between paired photons. However, quantum correlations can evolve over the lifetime of a multi-photon state. Here we demonstrate that states that show classical behaviour using standard measurements, in fact show entanglement when resolved as a function of time. Such entanglement could be efficiently utilized in quantum logic and security applications.

Entangled photon pairs can be generated by a number of techniques, including by parametric down conversion[3], CuCl crystals[4], two-photon interference[5], and quantum dots[6,7,8,9,10]. A single quantum dot emits a pair of photons as it decays radiatively from the initial biexction (XX) state, to the ground state (GS). The decay proceeds via one of two paths, which are represented in Figure 1(a) by the energy stored in the quantum dot as function of time. After a time $t_{XX}$ spent in the XX state, a biexciton photon $H_{XX}$ or $V_{XX}$ is emitted as the dot decays to the exciton (X) state. The polarisation of the biexciton photon is either horizontal (H) or vertical (V), and corresponds to decay into the exciton state $X_H$ or $X_V$ respectively. At this time ($t_{XX}$), the system exists in a symmetric superposition of the exciton-photon states $|H_{XX}X_H\rangle$ and $|V_{XX}X_V\rangle$. The quantum dot remains in a superposition of $X_H$ and $X_V$ for time delay $\tau$, during which a phase difference develops due to the energetic splitting $S$ between alternate exciton states. Subsequently, the exciton photon $H_X$ or $V_X$ is emitted with the same polarisation as that of the earlier biexciton photon, and the quantum dot reverts to the ground state. The system now exists as a superposition of orthogonally polarised photon pair states, with the phase between them determined by the time delay τ between the first (biexciton) and second (exciton) photon emission events. The final two-photon state is therefore $\Psi \propto \left( |H_{XX}H_X\rangle + e^{iS\tau/\hbar}|V_{XX}V_X\rangle \right) / \sqrt{2}$.

To illustrate the time dependent nature of the superposition, consider the biphoton (photon pair) intensity, which decays exponentially with delay τ, and the phase Φ, which evolves linearly with τ according to $\Phi = S\tau/\hbar$. The relationship between intensity and phase is shown using polar coordinates in Figure 1(b). For a quantum dot with zero splitting (red line), the superposition remains in phase as the intensity decays during an emission cycle. This results in the observation of a well-defined entangled photon pair state when integrated over delay τ. In contrast, for a quantum dot with finite splitting (black line), the phase of the superposition rotates as the intensity decays. Thus averaged over time, instantaneous superpositions largely cancel out with those at other times with opposing phase, giving rise to more classical photon pair states. This is the origin of the reduction in time-integrated entanglement as a function of splitting[11], and of the propensity to categorise the emission from quantum dots imparting spectral 'which-path' information as classical [12,13,14].

We will show that it is possible to resolve the hidden evolution of the entanglement properties as a function of the time delay τ. Only a few measurements of similar type have been reported previously, including evolution of entangled atom-photon systems[15,16]. However, in these experiments evolution is controlled using probe delay or other parameters, and the final state is an entangled two-photon state with fixed phase. Integrated over detection time, such states do not present classical behaviour, unlike those of a quantum dot. In other work, non-degenerate two-photon interference[17] showed strong maxima and minima resolved in time, despite poor interference averaged over time. Of course the situation with quantum dots is quite different, as it is interference between superpositions of exciton-photon pair states that drives evolution of entanglement.

The sample used was similar in design to those of previous experiments[8,9,11], and contains a single layer of InAs quantum dots, with dot density <1μm$^{-2}$. The dots are formed at the centre of a 1λ GaAs microcavity, defined by distributed Bragg reflectors consisting of 6 and 18 pairs of λ/4 AlAs/GaAs layers above and below respectively. Apertures of ~3μm diameter were fabricated in a metal film on the surface of the sample to isolate emission from individual dots. The sample was cooled to ~10K, and excited non-resonantly by a laser diode with ~100ps pulses at 80MHz. The splitting $S$ was controlled by applying an in-plane magnetic field[18], and determined by direct measurement of the polarisation dependent photoluminescence using a CCD camera[19].

To probe entanglement as a function of time, we measure the fidelity $f^+$ with the maximally entangled state $\Psi^+ = (|H_{XX}H_X\rangle + |V_{XX}V_X\rangle)/\sqrt{2}$, which is the expected state for a quantum dot with $S=0$[6,9]. The time parameter $\tau$ is the emission delay of the exciton (X) photon relative to the biexciton (XX) photon, and was selected by applying a single timing gate to accept only photon pairs for which $\tau_g \leq \tau \leq (\tau_g + w)$. This region is indicated on the predicted exponential decay of the biphoton intensity of Figure 2(a) by a yellow shaded region. The fidelity with $\Psi^+$ was determined using the relationship $f^+(\tau_g,w) = (C_R(\tau_g,w) + C_D(\tau_g,w) - C_C(\tau_g,w) + 1)/4$[11], which requires the measurement of the degree of correlation in the rectilinear ($C_R$), diagonal ($C_D$) and circular ($C_C$) polarisation bases. Here $C = (g^{(2)}_{\parallel} - g^{(2)}_{\perp})/(g^{(2)}_{\parallel} + g^{(2)}_{\perp})$, where co- and cross-polarised correlations $g^{(2)}_{\parallel}$ and $g^{(2)}_{\perp}$ are determined for an unpolarised source (as verified experimentally within 2% error). The second order cross correlation $g^{(2)}(\tau_g, w)$ was determined for different gate parameters by measuring $\tau$ for each photon pair using our previously reported APD based detection scheme. The error in $f^+$ is dominated by the poissonian counting statistics[20].

The fidelity $f^+$ of the emission from a dot with S=2.5±0.5μeV is plotted in Figure 2(b) as a function of the gate width $w$ as black points. The start of the gate is fixed at $\tau_g$=0, which we define as the modal delay between biexciton and exciton photon detection. For a gate width $w$=2ns, the fidelity $f^+$ is measured to be 0.46±0.01, which is below the 0.5 maximum achievable fidelity for an unpolarised classical state. However, as the gate width is reduced below ~1ns, the fidelity begins to increase, up to a maximum of 0.73±0.05 for the smallest gate width of $w$=49ps, indicating entanglement. This is a consequence of resolving entanglement before the state has significantly evolved over time.

We also plot the biphoton intensity measured within the gate, normalised to the total biphoton intensity for infinite $w$, as red points. The curve fits excellently to the predicted 1-exp(-$w/\tau_X$), revealing an exciton lifetime $\tau_X$ of 769±9ps. It is clear that large increases in fidelity can be achieved without a dramatic effect on the intensity of light collected.

We developed a model to calculate the expected behaviour. We begin by writing down the time dependent form of the biphoton density matrix $\underline{\underline{\rho}}$, derived using a time domain analysis of the intermediate entangled exciton-photon state[11].

$$\underline{\underline{\rho}}(\tau) = \frac{1}{4}\begin{pmatrix} 1+kg'^{(1)}_{H,V} & 0 & 0 & 2kg^{(1)}_{H,V}e^{-iS\tau/\hbar} \\ 0 & 1-kg'^{(1)}_{H,V} & 0 & 0 \\ 0 & 0 & 1-kg'^{(1)}_{H,V} & 0 \\ 2kg^{(1)}_{H,V}e^{iS\tau/\hbar} & 0 & 0 & 1+kg'^{(1)}_{H,V} \end{pmatrix}$$

Here, $g'^{(1)}_{H,V}$ is the fraction of dot emission unaffected by spin-scattering, $g^{(1)}_{H,V}$ the first-order cross-coherence, and $k$ the fraction of photon pairs that originate exclusively from the dot. All these parameters are in general time-dependent. For the fits presented here, we approximate the situation to the limit of no cross-dephasing ( $kg^{(1)}_{H,V} = kg'^{(1)}_{H,V}$ )[11]. The fidelity $f^+(\tau,w)$ is computed numerically using a Monte-Carlo approach, to incorporate a Gaussian approximation of the APD jitter observed in experiment.

The fidelity measured in the limit of large splitting is used to determine the time-integrated contribution from polarisation uncorrelated light ( $1-kg'^{(1)}_{H,V}$ )[11]. For simplicity, we approximate the fraction of uncorrelated light as time-independent. The same trend of increasing fidelity with reducing gate width $w$ is reproduced, as shown by the solid line in Figure 2(c).

The fidelity is increased for small gate widths because the system post-selects photons in the time-domain that have a similar phase relationship between the orthogonally polarised components of the superposition. In the measurements above, the choice of $\tau_g$=0 limits the phase acquired in the exciton state close to zero, so collected photons have high fidelity with the symmetric superposition $\Psi^+$. Similarly enhanced fidelities could be obtained for other values of $\tau_g$ with other maximally entangled states with different phase.

Selection in time equivalently reduces which-path information from the polarisation splitting $S$ in the energy domain. This is because the Fourier transform of a truncated exponential decay results in a broad natural linewidth of the post-selected photons. This is shown in Figure 2(d) by the Fourier transform of the biphoton decay, truncated after emission time delay $\tau$ of 0.39 or 1.0 ns.

In comparison to direct energy-resolved post-selection[21], resolving in time is more efficient. This is understandable as time-resolved post-selection targets photons at the beginning of the decay cycle, where emission intensity is strongest. In contrast, energy-resolved post-selection targets photons emitted with energies between those of $H_X$ and $V_X$, where intensity is minimum.

We note that the efficiency of the time selection technique can be increased further, giving rise to higher fidelity entanglement, whilst rejecting fewer biphotons. For example, by applying a second gate, delayed relative to the first to allow the phase in the exciton state to evolve a further $2\pi$, as shown schematically in Figure 2(a).

We measure next how the fidelity evolves over time. The gate width $w$ was fixed at $w$=537ps for $S$<4μeV, and 293ps for $S$>4μeV, in order to balance the requirements of low statistical error and significant fidelity improvement. Figure 3(a) plots the measured fidelity $f^+$ with ψ+ as a function of the delay $\tau_g$. Measurements for different splittings between $S=2.5\pm0.5$ and $13.5\pm0.5\mu eV$ (as marked) are offset vertically for clarity. Striking oscillations of the fidelity are observed, most clearly for the smallest investigated splitting of $S=2.5\mu eV$. The oscillatory behaviour is due to the phase of the superposed state rotating away from 0, and later returning to $2\pi$, which has maximum fidelity with $\psi+$. It is important to stress that when $f^+$ is minimum, entanglement still exists in the system but is expected to have high fidelity with the orthogonal state $\Psi^- = (|H_{XX}H_X\rangle - |V_{XX}V_X\rangle)/\sqrt{2}$.

The frequency of the oscillations increases as the splitting $S$ increases, accompanied by a reduction in the amplitude of the oscillations. The frequency range for which oscillations weaken is comparable to the measured timing jitter of FWHM=577ps introduced by photon pair detection using silicon avalanche photodiodes. We attribute the reduction in amplitude to time averaging of the oscillations as the frequency approaches the resolution limit of our system. For the same reason oscillations cannot be resolved for the largest $S$ measured of $13.5\mu eV$.

The calculations to reproduce the experimental results are shown in Figure 3(b). The solid lines represent the constant uncorrelated light fraction model described above, with the splitting and gate width parameters corresponding to those in the experiment. There is convincing agreement with the experimental data, and the trend of increasing frequency and reducing amplitude is reproduced well.

Improved agreement between experiment and theory in Figs. 2(c) and 3(b) is expected by incorporating non-trivial time dependence of parameters such as background light fraction, spin scattering time, and even polarisation splitting $S$. Further work is required to investigate these effects. However, a more realistic variant of the model can be constructed by including exponentially decaying background light, and a constant spin-scattering time. Estimation of the background light fraction accounts for most of the uncorrelated light observed, which suggests a spin-scattering time of ~8ns. The corresponding fits are shown as dashed lines in figures 2 and 3, and show similar agreement to the simple model.

The time integrated fraction of coherent dot light used in the model calculations is 0.78, which corresponds to a fidelity of 0.84 with the time evolving maximally entangled state $\Psi$. The experimental and model results are therefore consistent with the interpretation that the quantum dot always emits entangled light, even if the exciton level is not degenerate. The entangled state evolves as a function of $\tau$ which could be compensated for during measurement using knowledge of the splitting *S*.

Finally, an efficient way to utilise photons entangled in time-dependent states is to measure the detection times of both photons and to estimate the state, then feed back this information into the quantum information system. However, direct interactions between qubits entangled in time delay dependent states could reveal interesting physics, and might lead to radically different implementations of quantum logic.

In conclusion, we have shown that quantum dots with non-zero polarisation splitting emit photon pairs into a time-evolving entangled state. Such entanglement is hidden from conventional time-integrated measurement techniques, which previously led to the belief that such dots generate only classical light. The fidelity of entanglement is found to be comparable to dots with degenerate exciton states, and could be utilised in applications adapted for evolving states. Our research highlights that by selecting the correct measurement approach, entanglement in different, but useful, forms can be extracted from seemingly classical sources.

The authors gratefully acknowledge financial support from QIP IRC, EC FP6 Network of Excellence, Sandie, QAP, and the EPSRC.

[21] N. Akopian et al., *Phys. Rev. Lett* **96**, 130501 (2006).

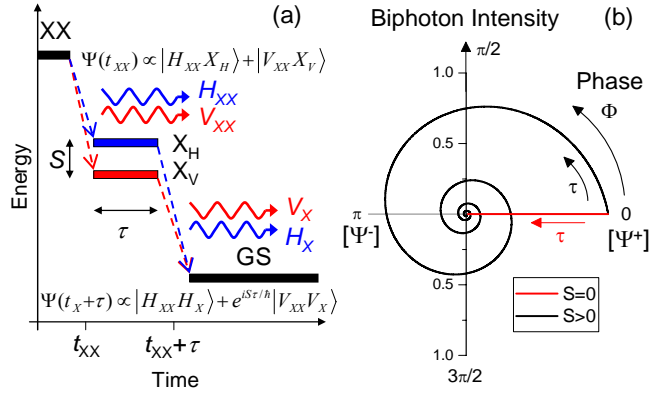

Figure 1. Schematics of entangled photon pair generation in quantum dots. (a) shows energy of a quantum dot as a function of time following excitation to the initial biexciton (XX) state. The state ψ is marked for times corresponding to emission of the first and second photons. (b) represents the relationship between the biphoton intensity and the phase of the photon pair superposition for dots with zero (red) and finite (black) fine structure splitting S.

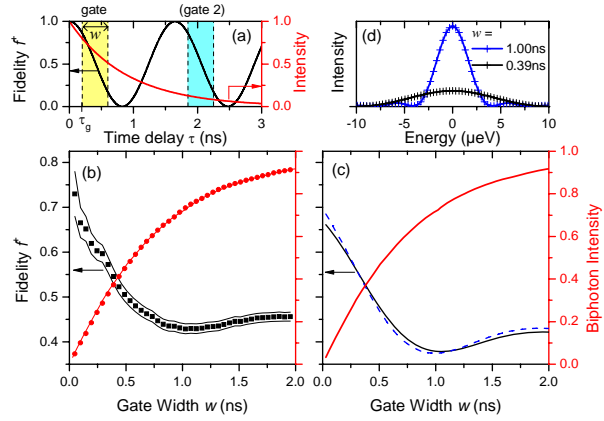

Figure 2. Recovery of entanglement by time discrimination. (a) Schematic of the time-dependent gate(s) applied to post-select photon pairs based on the time interval τ. Black line represents the fidelity (left axis), and the red line represents the photon pair intensity (right axis). (b) Fidelity $f^+$ and fraction of photon pairs retained after post-selection in time within a gate of width *w*, begging at $\tau_g$=0. (c) Calculated behaviour corresponding to (b). (d) Measured natural linewidth of photon pairs post-selected with a 1ns (blue) and 0.39ns (black) single gate.

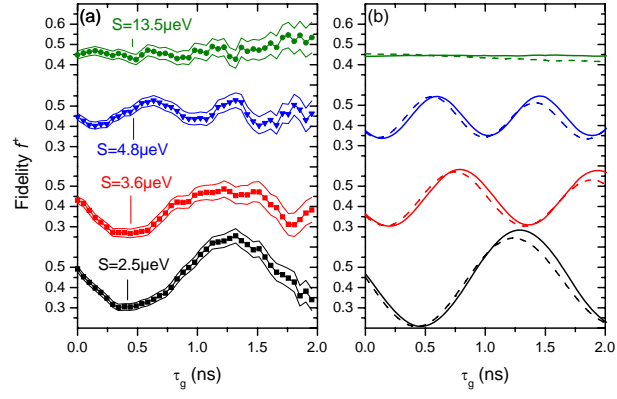

Figure 3. Experimental and calculated fidelity $f$ as a function of the time between photons $\tau$. (a) Measured fidelity for a single quantum dot with different fine structure splitting S as indicated. Bars denote poissonian counting errors. (b) Calculated fidelity corresponding to experimental conditions in (a).